Examining the Wigner Distribution Using Dirac Notation

Frank Rioux
Department of Chemistry
St. John's University and College of St. Benedict
St. Joseph, MN 56374

Expressing the Wigner distribution function in Dirac notation reveals its resemblance to a classical trajectory in phase space.

The purpose of this note is to demonstrate that expressing the Wigner distribution [1–3] in Dirac notation demonstrates its resemblance to a classical phase-space trajectory. The Wigner distribution can be generated from either the coordinate- or momentum-space wave function. The Wigner transform using the coordinate-space wave function is given in equation (1) for a one-dimensional example in atomic units.

$$W(p,x) = \frac{1}{2\pi}\int_{-\infty}^{\infty} \Psi^*(x+\tfrac{s}{2})\Psi(x-\tfrac{s}{2})e^{ips}ds \qquad \textbf{(1)}$$

In Dirac notation the first two terms of the integrand are written as follows,

$$\Psi^*(x+\tfrac{s}{2}) = \langle \Psi | x+\tfrac{s}{2}\rangle \qquad \Psi(x-\tfrac{s}{2}) = \langle x-\tfrac{s}{2} | \Psi\rangle \qquad \textbf{(2)}$$

Assigning 1/2π to the third term and utilizing the momentum eigenfunction in coordinate space and its complex conjugate we have,

$$\tfrac{1}{2\pi}\mathrm{e}^{ips} = \tfrac{1}{\sqrt{2\pi}}\mathrm{e}^{ip(x+\frac{s}{2})}\,\tfrac{1}{\sqrt{2\pi}}\mathrm{e}^{-ip(x-\frac{s}{2})} = \langle x+\tfrac{s}{2}|p\rangle\langle p|x-\tfrac{s}{2}\rangle \qquad \textbf{(3)}$$

Substituting equations (2) and (3) into equation (1) yields after rearrangement,

$$W(x,p) = \int_{-\infty}^{+\infty}\langle\Psi|x+\tfrac{s}{2}\rangle\langle x+\tfrac{s}{2}|p\rangle\langle p|x-\tfrac{s}{2}\rangle\langle x-\tfrac{s}{2}|\Psi\rangle ds \qquad \textbf{(4)}$$

The four Dirac brackets are read from right to left as follows: (1) is the amplitude that a particle in the state Ψ has position (*x - s/2*); (2) is the amplitude that a particle with position (*x - s/2*) has momentum $p$; (3) is the amplitude that a particle with momentum $p$ has position $(x + s/2)$; (4) is the amplitude that a particle with position $(x + s/2)$ is (still) in the state $\Psi$. Thus, in Dirac notation the integrand is the quantum equivalent of a classical phase-space trajectory for a quantum system in the state $\Psi$.

Integration over $\boldsymbol{s}$ creates a superposition of all possible quantum trajectories of the state $\Psi$, which interfere constructively and destructively, providing a quasi-probability distribution in phase space. As an example, the Wigner probability distribution for the double-slit experiment is shown in the figure below [4, 5]. The oscillating positive and negative values in the middle of the Wigner distribution signify the interference associated with a quantum superposition, distinguishing it from a classical phase-space probability distribution. In the words of Leibfried et al. [5], the Wigner function is a "mathematical construct for visualizing quantum trajectories in phase space."

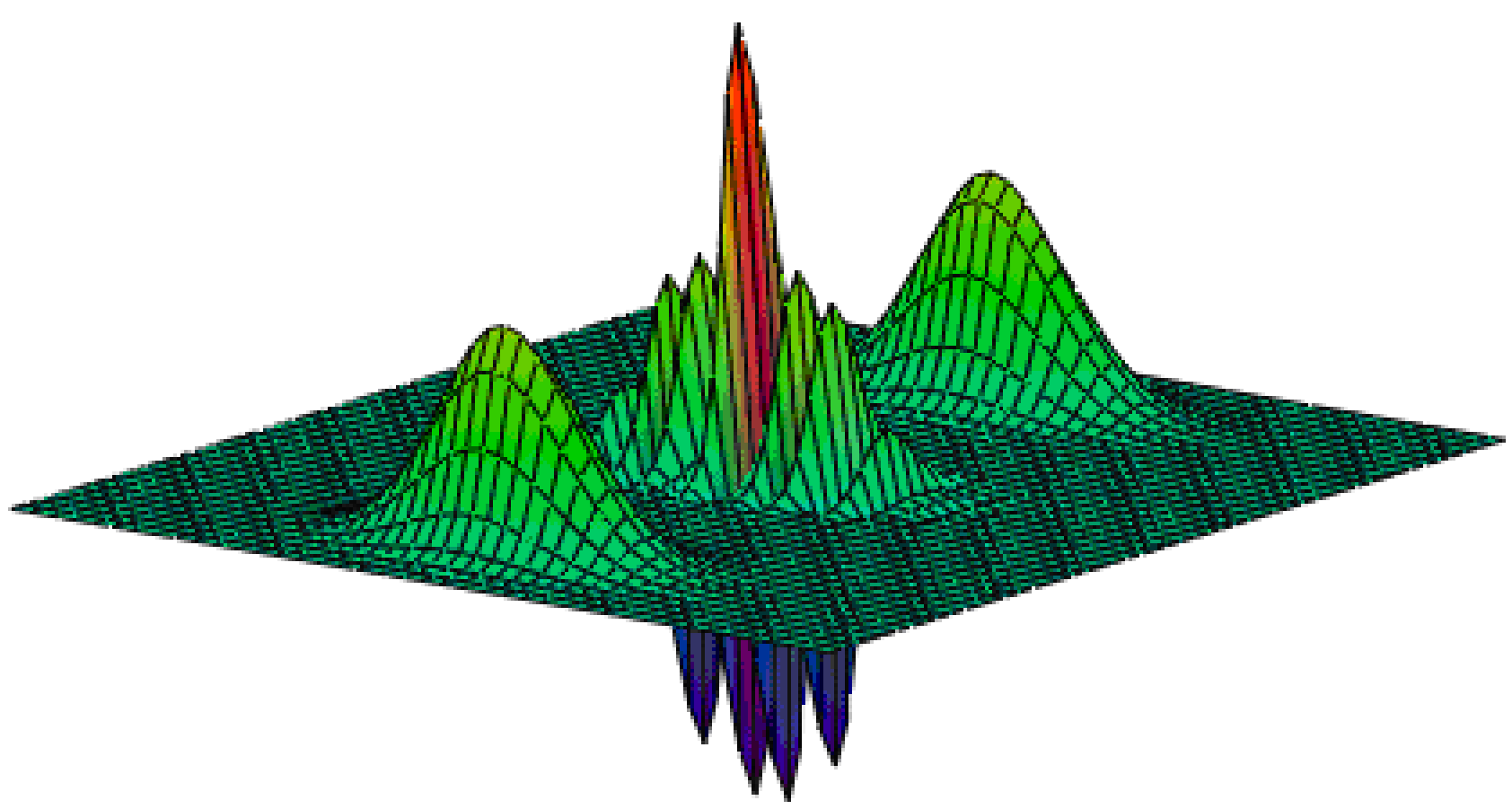

Wigner distribution function for the double-slit experiment.

**Other relevant literature references:**